# Hidden order in online extremism and its disruption by nudging collective chemistry

N.F. Johnson[1,2], N. Velásquez[2], P. Manrique[3], R. Sear[4], R. Leahy[2,5], N. Johnson Restrepo[2,5], L. Illari[1], Y. Lupu[6]
[1]Physics Department, George Washington University, Washington D.C. 20052
[2]Institute for Data, Democracy and Politics, George Washington University, Washington D.C. 20052
[3]Theoretical Biology and Biophysics Group, Los Alamos National Laboratory, Los Alamos, New Mexico, 87545
[4]Department of Computer Science, George Washington University, Washington D.C. 20052
[5]ClustrX LLC, Washington D.C.
[6]Department of Political Science, George Washington University, Washington D.C. 20052

**We show that the eclectic "Boogaloo" extremist movement that is now rising to prominence in the U.S., has a hidden online mathematical order that is identical to ISIS during its early development, despite their stark ideological, geographical and cultural differences. The evolution of each across scales follows a single shockwave equation that accounts for individual heterogeneity in online interactions. This equation predicts how to disrupt the onset and 'flatten the curve' of such online extremism by nudging its collective chemistry.**

Disrupting the emergence and evolution of potentially violent extremist movements is a crucial challenge. In recent months, Facebook has designated the new U.S. Boogaloo movement a violent anti-government network and dangerous organization[1], U.S. Congress has been alerted to potential Boogaloo violence[2], and a Boogaloo member has been arrested for the death of a federal officer[3]. The Boogaloo movement came to prominence in the U.S. through Facebook in 2020, with a highly diverse mix of narratives ranging from Second Amendment gun rights and Black Lives Matter racism through to COVID-19 lockdown protests and upcoming U.S. elections, with members drawn from across conservative, libertarian and nihilistic ideologies. By contrast, ISIS (Islamic State) is a pro-jihad, anti-U.S. movement whose significant growth in 2015 developed outside the U.S. on central European platforms such as VKontakte (see Supplementary Material SM). These movements vary dramatically in terms of their evolution, ideology, goals, geography and drivers such as political grievances, poverty and personality traits[4,5,6]. Existing extremism research has addressed such features in detail with great success[4,5,6,7,8,9,10,11,12,13,14,15,16]. Our focus provides a complementary, system-level understanding that yields fresh quantitative insight into their evolution and possible interventions.

Given these many features and differences, it is not expected that the Boogaloo and ISIS movements would possess any deep mathematical order or that it would be shared by both. Indeed, each evolves very differently over time, both internally (Fig. 1A,B) and as a whole (Fig. 2). However, we find that the emergence and evolution of each follows the predictions of a single shockwave equation, both at the level of the individual communities within each movement (Fig. 1A,B) and across movements (Fig. 2). The key feature of our mathematical theory, which we illustrate in Fig. 1C and derive in the SM, is that it incorporates individual human heterogeneity into the online aggregation process. The heterogeneity of each individual $i$ is mimicked by a vector $\vec{x}_i$ which can be of arbitrary complexity (i.e. any number of dimensions) and can in principle change over time. This results in heterogeneity-dependent interactions between individuals which, when averaged over all pairs, gives a value $F$. How online communities emerge and evolve, then depends on this collective online 'chemistry'.

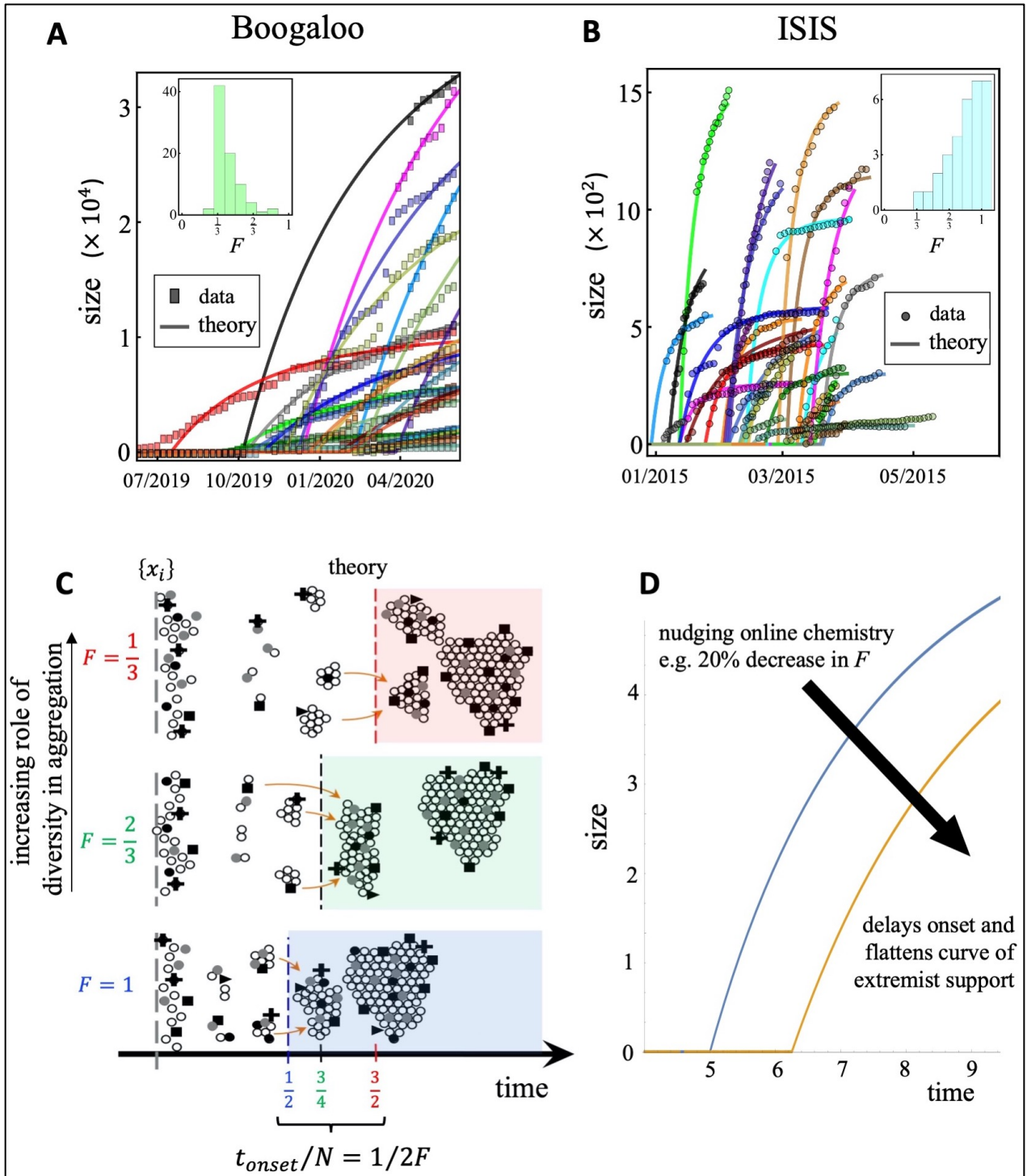

**Fig. 1. A and B:** Growth curves of online Boogaloo communities (each is a Facebook Page whose size is the number of members) and online ISIS communities (each is a VKontakte Group whose size is the number of members). Their empirical growths (symbols) differ and occur over different timescales, yet they follow the same mathematical equation that incorporates individual user heterogeneity into the online social aggregation process (solid lines). For visual clarity, only a few of the communities are shown in the main plots. Our data collection follows previous work[17,18,19,20,21] and focuses on communities (Pages, Groups) since these play a greater role in nurturing narratives than platforms like Twitter which have no pre-built community tool. **C:** Schematic of our mathematical theory showing predicted onset times for example $F$ values (see SM). **D:** Predicted curves from our theory (arbitrary units) show how onset can be delayed and the curve flattened by nudging $F$ and hence nudging the collective chemistry.

Our mathematical theory of aggregation-with-heterogeneity (Fig. 1C) makes specific predictions:
(1) There is a single equation that governs the emergence and evolution of each individual extremist community (Fig. 1A,B) and also an entire extremist movement (Fig. 2A,B). It is a generalized shockwave equation $\frac{\delta \mathcal{E}}{\delta t} = \frac{\delta \mathcal{E}}{\delta y} \frac{2F}{N} \left( \frac{\mathcal{E}}{N} - 1 \right)$ where $F$ and $N$ measure the heterogeneity and user pool size for individual communities (Fig. 1) or an entire movement (Fig. 2), and $\mathcal{E} = \sum_{k=1}^{N} k n_k e^{yk}$.
(2) The size of each community (Fig. 1) and the entire movement (Fig. 2) is predicted to vary in time as $G(t) = N \left( 1 - W \left( \left[ \frac{-2Ft}{N} \right] exp \left[ \frac{-2Ft}{N} \right] \right) / \left[ \frac{-2Ft}{N} \right] \right)$ where $W$ is the Lambert function.
(3) Each community in Fig. 1 and the entire movement in Fig. 2, is predicted to have its own tipping point time $t_{onset} = \frac{N}{2F}$ which signals the onset of a macroscopic swell in online support. This tipping point is, in the limit of large $N$, a dynamical phase transition.
(4) The community size distribution at the onset $t_{onset}$ (see Fig. 2 insets) is predicted to be a power-law with exponent $-5/2 = -2.5$.

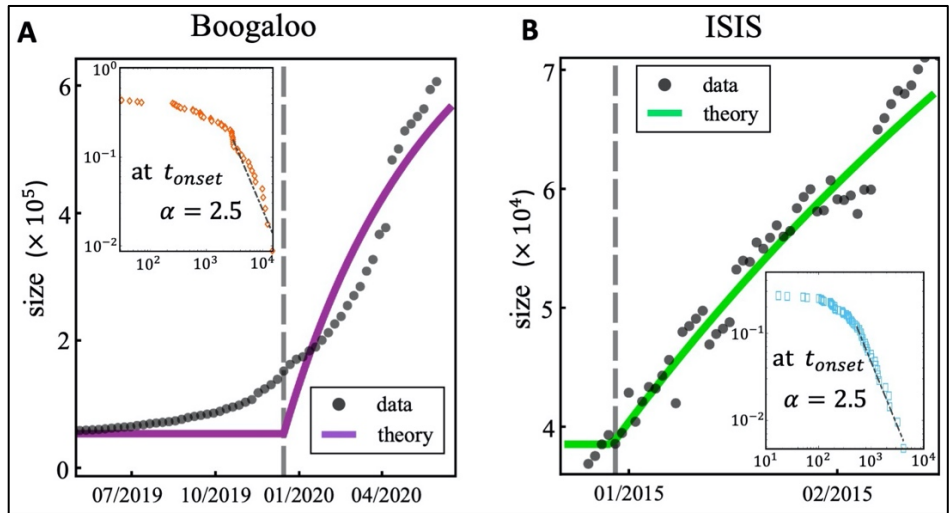

**Fig. 2: A and B: Growth curves of the entire Boogaloo movement (size is the combined number of members of their Facebook Pages), and the entire ISIS movement (size is the combined number of members of their VKontakte Groups). Insets show distribution of sizes of the individual communities at the predicted onset $t_{onset}$ (vertical gray line in main panels). The maximum-likelihood estimate for the power-law exponent in each case is $-2.5$ to two significant figures, which is the same value as predicted by the theory (SM).**

The onset times $t_{onset}$ and growth curves $G(t)$ for each Boogaloo community and ISIS community in Figs. 1A,B, and for the movements as a whole in Fig. 2A,B, are well described by these mathematical predictions. Moreover, Fig. 2A,B insets show that the individual community sizes at $t_{onset}$ exhibit the predicted power-law exponent $-2.5$. Since the theoretical formulae are derived for $N$ very large, the predicted onsets are too sharp: but at the expense of losing the closed-form formulae, we can extend the theory to account for finite $N$ as in the empirical data. The transition then becomes smooth like the empirical data, with the size at the onset varying as $[N]^{-1/3}$ as opposed to being strictly zero.

The inferred membership heterogeneity is similar for each movement (Figs. 2A,B) with both $F$ values being statistically indistinguishable from $1/3 = 0.33$, which is the value predicted mathematically for aggregation favoring diversity in a population with uniformly distributed $\{\vec{x}_i\}$ (see SM). This would suggest that each movement develops by aggregating diverse sets of supporters from the global online user pool, instead of some particular pre-polarized population. The Fig. 1A inset shows that the membership heterogeneities $F$ of individual Boogaloo communities are also close to $1/3$ which suggests that individual Boogaloo community formation is also driven by the same preference for diversity as the entire Boogaloo movement. This is consistent with the eclectic mix of memes and ideas that we observe in the content of each Boogaloo community, and the lack of any increase in topic coherence that we observe from our dynamic Latent Dirichlet Allocation topic analysis of their narratives (see SM). By contrast, individual ISIS communities have $F$ values closer to 2/3 (Fig. 1B inset) which suggests that once inside the ISIS movement, its supporters form into communities which are each internally homogenous and have a well-defined narrative flavor. Overall, this suggests that while Boogaloo and ISIS recruits join the overall movements driven by diversity, Boogaloos continue with this diversity driver when forming and joining an individual community, while ISIS supporters prefer a community to have a single narrative flavor.

By incorporating the interplay between individual human heterogeneity and connective action[22] online (Figs. 1, 2), our mathematical theory has placed these extremist movements' evolutions on the same footing -- akin to a single equation in physics explaining the different trajectories of a plane taking off vs. landing in terms of uplift vs. gravity. While there will always be different trajectories for different movements and different communities within each movement, this identification of a shared mathematical order[23,24,25,26,27] opens the door to a common set of mitigation strategies.

An immediate example is the mathematical theory's prediction that online extremism can be mitigated by nudging the online collective chemistry. Slightly decreasing $F$ at the level of the entire movement or individual communities (Fig. 1D) delays the onset (since $t_{onset} \propto 1/F$) and flattens the growth curve $G(t)$: specifically, $-\Delta F/F = \Delta t_{onset}/t_{onset}$. This could be achieved by biasing the interaction of dissimilar individuals such that aggregation favors dissimilar $\vec{x}_i$'s, by injecting their online space (e.g. Facebook Page) with more diverse material, or by nudging the composition of the overall pool of potential recruits (see SM). Alternatively, by nudging smaller (and hence potentially less robust) communities, the power-law distribution at the onset can be disrupted (see SM) which in turn disrupts the dynamical phase transition and hence delays the onset of support. These approaches, while significant challenges in practice, are preferable to blanket shutting down of all communities: the latter would rely on them all being found, risks claims of freedom-of-speech, and could ignite new support.

Explaining *why* the Boogaloos have suddenly emerged requires deeper social, political and economic debate. However, our mathematical analysis offers a quantitative answer by extending the heterogeneity-driven aggregation toward a generalized version of the sociological Seceder Model of Dittrich et al. and Halpin-Healy et al.[28,29,30] (Fig. 3, SM) which features a competition between the pressure to conform and the desire to dissent. Such a competition is consistent with the Boogaloos' eclectic mix of fads (e.g. memes) and fashions and the lack of any increasing topic coherence in their narratives (see SM). This model correctly predicts the emergence of 3 stable movements (Fig. 3) and it could even be used to estimate the number of new extremist 'branches' to expect in the future.

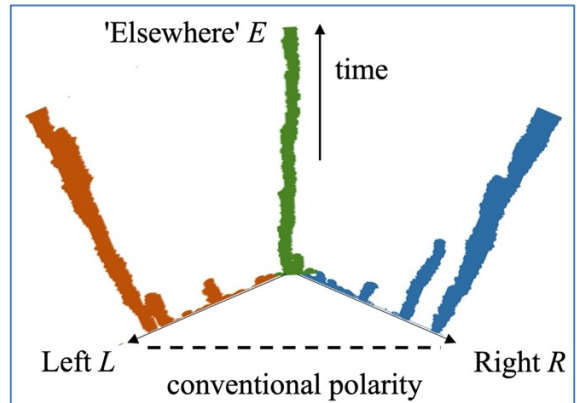

**Fig. 3.** Computer simulation predicts Boogaloo emergence outside the usual Left-Right spectrum. It is generated by making the aggregation in Fig. 1C follow the Seceder mechanism of Dittrich et al. and Halpin-Healy et al.[28,29,30]. We show this in a hybrid two-dimensional space since the emergence and stability are general for many model variants, including the simpler case of a one-dimensional $\vec{x}_i$ which we used to generate the branches shown.

Among potential limitations of our study, is the fact that our mathematical analysis will always be an approximation and hence the theoretical predictions in Figs. 1 and 2 are not perfect. While the smooth onset of the empirical growth curves can be reproduced at the expense of a loss of closed-form formulae as discussed earlier, there are still unexplained bumps and jumps. However, these can also be reproduced if we allow for an online influx of potential recruits and hence $N$ becomes a function of time. We also need to study future extremist movements as they emerge over time online, to check the generality of our findings. However, the Boogaloos and ISIS are certainly movements of high current importance -- and the SM shows that their hidden mathematical order reported here, does not arise for other online human aggregation behaviors. Finally, we stress that our mathematical analysis is targeted at complementing current analyses in the social sciences, not to replace them.